%% file: DPPProblemFormulationv3.tex
\documentclass[final]{IEEEtran}

\usepackage{amsthm,amssymb,graphicx,multirow,amsmath,color,amsfonts}
\usepackage[update,prepend]{epstopdf}
\usepackage[noadjust]{cite}
\usepackage[latin1]{inputenc}
\usepackage{tikz}
\usetikzlibrary{arrows,calc}		
\usepackage{bbm} 
\usepackage{pdfpages}
\usepackage{tabulary}
\usepackage{multirow}
\usepackage{comment}
\usepackage{subfigure,array}
\usepackage{bm}
\usepackage{float}

\allowdisplaybreaks
\usepackage{algorithm}
\usepackage[noend]{algpseudocode}

\makeatletter
\def\BState{\State\hskip-\ALG@thistlm}
\makeatother

\title{Machine Learning meets Stochastic Geometry:\\ Determinantal Subset Selection for Wireless Networks}
\author{
Chiranjib Saha and Harpreet S. Dhillon 
\thanks{The authors are with Wireless@VT, Department of ECE, Virginia Tech, Blacksburg, VA, USA. Email: \{csaha,   hdhillon\}@vt.edu. The support of the US National Science Foundation (Grant CNS-1617896) is gratefully acknowledged. 
} \vspace{-0.8cm}}
\include{notationITA}

\begin{document}
\maketitle
\vspace{-21cm}
\begin{abstract}
In wireless networks, many problems can be formulated as {\em subset selection problems} where the goal is to select a subset from the ground set with the objective of maximizing some objective function. These problems are typically NP-hard and hence solved through carefully constructed heuristics, which are themselves mostly NP-complete and thus not easily applicable to large networks.  On the other hand,  subset selection problems occur in slightly different context in machine learning (ML)    where the goal is to select a subset of high  {\em quality} yet {\em diverse} items  from a ground set. 
This balance in quality and diversity is often maintained in the ML problems by using determinantal point process (DPP), which endows distributions on the subsets  such that the probability of selecting two {\em similar} items is negatively correlated. While DPPs have been explored more generally in stochastic geometry (SG) to model inter-point repulsion, they are particularly conducive for ML applications because the parameters of their distributions can be efficiently learnt from a training set. 
In this paper, we introduce a novel DPP-based learning (DPPL) framework  for efficiently solving subset selection problems in wireless networks.  The DPPL is intended to replace the traditional optimization algorithms for subset selection by {\em learning} the quality-diversity trade-off  in the optimal subsets selected by an optimization routine.   As a case study,  we apply DPPL to the wireless link scheduling problem, where the goal is to determine the subset  of  simultaneously active links which maximizes the network-wide sum-rate. We demonstrate that the proposed DPPL approaches the optimal solution with significantly lower computational complexity than the popular optimization algorithms used for this problem in the literature. 
\end{abstract}
\section{Introduction}\label{sec::intro}
\vspace{.1cm}
 ML and SG have recently found many applications in the design and analysis of wireless networks. However, since the nature of the problems studied with these tools are so fundamentally different, it is  rare to find a common ground where the strength of these tools   can be jointly leveraged.  While the foundation of wireless networks  is built on  traditional  probabilistic models (such as channel, noise, interference, queuing models), ML is changing this {\em model-driven}  approach to more {\em data-driven} simulation-based approach by  {\em learning} the models from extensive datasets  available from the real networks or field trials~\cite{simeone2018very}. On the other hand, the basic premise of  SG is to enhance the {\em model-driven} approach by endowing distributions on the locations of the transmitters (Tx-s) and receivers (Rx-s) so that one can derive  the exact and tractable expressions for key performance metrics such as interference, coverage, and rate. In this paper, we concretely demonstrate that these two mathematical tools can be jointly applied to a class of problems known on the {\em subset selection problems}, which have numerous  applications in wireless networks. 

{\em Subset selection problems.} 
In wireless networks, a
 wide class of resource management problems like power/rate control, link scheduling, network utility maximization, and  beamformer design  fall into the category of {\em subset selection problems} where a subset from a ground set needs to be chosen which optimizes a given objective function. For most of the cases, finding the optimal subset is  NP-hard. The common practice in the  literature is to design  some heuristic algorithms which find a local optimum under reasonable  complexity. Even most of these heuristic approaches are NP-complete and are hence difficult to implement when the network size grows large. 

 In ML, subset selection problems appear in a slightly different context  where  the   primary objective is to preserve the balance between {\em quality} and {\em diversity} of the items in the subset, i.e., to select  good quality items from a ground set which are also non-overlapping in terms of their features.   
 For example, assume that a user is searching the images of New York in a web-browser. The image search engine will pick a subset of stock images related to New York from the image library which contains the popular landmarks ({\em quality}) as well as ensure that one particular landmark does not occur repeatedly the search result ({\em diversity}).  
 Few more examples  of  subset selection with diversity are  text summarization~\cite{nenkova2006compositional}, citation management~\cite{kulesza2012determinantal}, and sensor placement~\cite{krause2008near}.  The attempt to model diversity among the items in a subset selection problem brings us to the probabilistic models constructed by DPPs,  which lie at  the  intersection of ML and SG. Initially formulated as a repulsive point process in SG~\cite{li2015statistical}, DPPs are natural choice for inducing diversity or negative correlation between the items in a subset.   Although the traditional theoritical  development  of DPPs has been focused on the continuous spaces, 
 the finite version of the DPPs have recently emerged as  useful probabilistic models for the subset selection problems with {\em quality-diversity trade-off} in ML. This is due to  the fact that the finite DPPs are amenable to the data-driven learning and inference framework of ML~\cite{kulesza2012determinantal}. 
 
 {\em Relevant prior art on DPPs. }
 In wireless networks, DPPs have mostly been used  in the SG-based modeling and analysis of cellular networks. In these models, DPPs are used to capture spatial repulsion in the BS locations, which cannot be modeled using more popular Poisson point process (PPP)~\cite{li2015statistical}.  For some specific DPPs, for instance the Ginibre point process, it is possible to analytically  characterize the performance  metrics of the network such as the coverage probability~\cite{miyoshi2014cellular}. However, the finite DPPs and the associated data-driven learning framework, which is under rapid development in the ML community has not found any notable  application in wireless networks. The only existing work is   \cite{blaszczyszyn2018determinantal}, where the authors have introduced a new class of {\em data-driven SG models} using DPP and have trained them to mimic the properties of some hard-core point processes used for wireless network modeling  (such as the \matern type-II process) in a finite window. 
 
{\em Contributions.} 
The key technical contribution of this paper is the novel {\em DPPL framework} for solving general subset selection problems in wireless networks. 
 In order to concretely demonstrate the proposed DPPL framework, we   apply it to solve the link scheduling problem which is a classical  subset selection problem in wireless networks. 
The objective is to assign optimal binary power levels to Tx-Rx pairs so as to maximize the sum-rate~\cite{weeraddana2012weighted}. The links transmitting at a higher (lower) power level will be termed active (inactive) links. Therefore, the objective is to determine the optimal subset of {\em simultaneously active links}. Similar to the  subset selection problems in ML, the simultaneously active links will be selected by balancing between the quality and diversity.  The links which will be naturally favored are the ones with better link quality in terms of signal-to-interference-and-noise-ratio ($\sinr$) so that the rates on these links contribute more to  the sum-rate ({\em quality}). On the other hand, the simultaneously active links will have some degree of spatial repulsion to avoid mutual  interference ({\em diversity}).  With this insight, it is reasonable to treat the set of active links in the optimal
solution as a DPP over the set of links in a given network.  The DPP is trained by a sequence of networks  and their optimal subsets which  are generated by using an optimization algorithm based on geometric programming (GP). We observe that  the sum-rates of the estimated optimal subsets generated by the trained DPP  closely approach the optimal sum-rates.    Moreover, we show that the subset selection using DPP is significantly more  computationally efficient than the optimization based subset selection methods. 
\section{Determinantal point process: Preliminaries}
\begin{figure}
\centering
\subfigure[$g({\bf a}_1)$ increases.]{\includegraphics[scale=0.5]{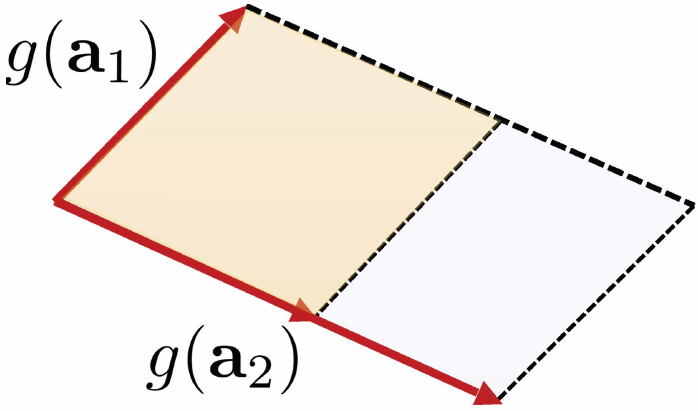}}\hspace{0.8em}
\subfigure[$S_{1,2}$ increases.]{\includegraphics[scale=0.5]{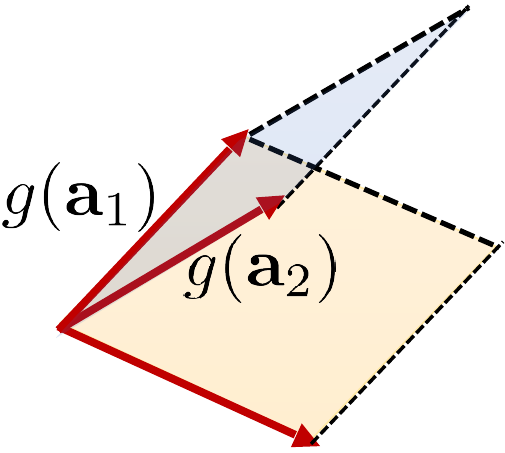}}
\caption{In DPP, the probability of occurrence of a set $Y$ depends on the volume of the parallelopiped with sides $g({\bf a}_i)$ and angles proportional to $\arccos(S_{i,j})$: (a) as $g({\bf a}_i)$ increases, the volume increases, (b) as $S_{i,j}$ increases, the volume decreases.}\label{fig::DPP::inside}
\end{figure}

\vspace{.7em}
In this Section, we provide a concise introduction to DPP on finite sets. The interested readers may refer to \cite{kulesza2012determinantal} for a more pedagogical treatment of the
topic as well as extensive surveys of the prior art. In general, DPPs are probabilistic models that quantify the
likelihood of selecting a subset of items as the determinant of a kernel matrix ($K$). More formally, if ${\cal Y} = \{1,\dots,N\}$ is a discrete set of $N$ items, a DPP  is a probability measure on the power set $2^{\cal Y}$ which is defined as:
\begin{equation}\label{eq::DPP::def::K::matrix}
\scalebox{0.9}{$
{\cal P}(A\subseteq {\bf Y}) = {\rm det}(K_A),$}
\end{equation}  
where  ${\bf Y}\sim {\cal P}$ is a random subset of ${\cal Y}$ and $K_A \equiv [K_{i,j}]_{i,j\in A}$ denotes the restriction on $K\in \R^{N\times N}$ to the indices of the elements of $A\subseteq {\cal Y}$ ($K_{\emptyset}=1$). We denote  $K$   as  the marginal kernel which is a positive semidefinite matrix such that $K\preceq I$ ($I$ is an $N\times N$ identity matrix), i.e. all eigenvalues of $K$ are less than or equal to 1.  For the learning purposes, it is more useful to define DPP with another formalism known as the {\em $L$-ensemble}. A DPP can be alternatively defined in terms of a matrix $L$ ($L\preceq I$) indexed by ${Y}\subseteq {\cal Y}$:
\begin{equation}\label{eq::DPP::def::L-ensemble}
\scalebox{0.9}{$
{\cal P}_L(Y) \equiv  {\cal P}_L({\bf Y} = Y  ) =\frac{{\rm det}(L_Y)}{\sum_{Y'\in  2^{\cal Y}}{\rm det}(L_{Y'})} = \frac{{\rm det}(L_Y)}{{\rm det}(L+I)},$} 
\end{equation}
 where $L_Y = [L_{i,j}]_{{i,j}\in Y}$. The last step follows from the identity $\sum_{Y'\in 2^{\cal Y}}{\rm det}(L_{Y'}) = {\rm det}(L+I)$ (see \cite[Theorem 2.1]{kulesza2012determinantal} for proof). 
Following \cite[Theorem 2.2]{kulesza2012determinantal},  $K$ and $L$  are related by the following equation:
\begin{equation}\label{eq::K-L::connection}
K = (L+I)^{-1}L . 
\end{equation} 
Since, $L$ is real and symmetric by definition, its eigendecomposition is $L = \sum_{n=1}^N \lambda_n {\bf v}_n{\bf v}_n^{\top}$, where $\{{\bf v}_n\}$ is the orthonormal sequence of eigenvectors corresponding to the eigenvalues $\{\lambda_n\}$. Using \eqref{eq::K-L::connection},   $K$ can also be obtained  by rescaling the eigenvalues of $L$ as:
\begin{equation}\label{eq::K::L::eigen}
\scalebox{1.0}{
$K  = \sum_{n=1}^N \frac{\lambda_n }{1+\lambda_n}{\bf v}_n{\bf v}_n^{\top}.$}
\end{equation}
In the ML formalism, if ${\bf a}_i\in \R^N$ is some vector representation of the $i^{th}$ item of ${\cal Y}$, then $L\in\R^{N\times N}$ can be interpreted as a kernel matrix, i.e., $L_{i,j} = k({\bf a}_i,{\bf a}_j)\equiv \phi({\bf a}_i)^{\top}\phi({\bf a}_j),$ where $k(\cdot,\cdot)$ is a kernel function and $\phi$ is the corresponding feature map. The kernel $k({\bf a}_i,{\bf a}_j)$ can be further decomposed according to the quality-diversity decomposition~\cite{kulesza2012determinantal} as: 
\begin{equation}\label{eq::kernel::decompostion}
L_{i,j} = k({\bf a}_i,{\bf a}_j) = g({\bf a}_i) S_{i,j} g({\bf a}_j),
\end{equation} where $g({\bf a}_i)$ denotes the quality of ${\bf a}_i$ ($\forall i\in{\cal Y}$) and $S_{i,j}= L_{i,j}/\sqrt{L_{i,i}L_{j,j}}$ denotes the similarity of ${\bf a}_i$ and ${\bf a}_j$ ($\forall i,j\in{\cal Y},i\neq j$).  Using~\eqref{eq::kernel::decompostion}, we can write \eqref{eq::DPP::def::L-ensemble} after some manipulation as:
 ${\cal P}_L ({\bf Y}=Y) \propto {\rm det}(L_Y) = {\rm det}(S_Y)\prod\limits_{i\in Y}g({\bf a}_i)^2,$
where the first term denotes the diversity and second term denotes the quality of the items in $Y$. We now provide a geometric interpretation of ${\cal P}_L ({\bf Y}=Y) $ as follows.
\begin{remark}
We can intuitively interpret  ${\rm det}(L_{ Y })$  as the   squared volume of the parallelepiped  spanned  by the vectors $\{\phi({\bf a}_i)\}_{i\in {\cal Y}}$, 
 where $\|\phi({\bf a}_i)\| = g({\bf a}_i)$ and $\angle\{\phi({\bf a}_i),\phi({\bf a}_j)\} = \arccos(S_{i,j})$. Thus,  items with higher 
 $g({\bf a}_i)$ are more probable since the corresponding $\phi({\bf a}_i)$-s span larger volumes. Also diverse items are more probable than the similar items since  more orthogonal collection of $\phi({\bf a}_i)$-s span larger volume (see Fig.~\ref{fig::DPP::inside} for an illustration). 
  Thus DPP naturally balances the  quality and diversity of items in a subset.
\end{remark}
\section{The Proposed DPPL framework}
\subsection{Conditional DPPs}
Most of the learning applications are {\em input-driven}. For instance, recalling the image search example, a user input will be required to show the search results. To model these input-driven  problems,  we require  conditional DPPs. In this framework, let $X$ be an external input. 
Let ${\cal Y}(X)$ be the collection of all possible candidate subsets given $X$.
 The conditional DPP assigns {probability to every possible subset} $Y\subseteq {\cal Y}(X)$ as:
${\cal P}({\bf Y} = Y|X)\propto {\rm det}(L_{Y}(X))$, 
where $L_{Y}(X)\in (\R^+)^{|{\cal Y}(X)|\times |{\cal Y}(X)|}$ is a positive semidefinite kernel matrix. Following \eqref{eq::DPP::def::L-ensemble}, the normalization constant  is   ${\rm det}(I+L(X))$. Now, similar to the decomposition technique in \eqref{eq::kernel::decompostion},  
 $
L_{i,j}(X) = g({\bf a}_i|X)S_{i,j}(X) g({\bf a}_j|X),
$ 
where  $g({\bf a}_i|X)$ denotes the quality measure of link $i$ and $S_{i,j}(X)$ denotes the diversity measure of the links $i$ and $j$ ($i\neq j$) given $X$. In \cite{kulesza2012determinantal}, the authors proposed  a log-linear model  for the quality measure as follows:
\begin{equation}\label{eq::model::quality}
g({\bf a}_i|X) = \exp\left(\bm{\theta}^{\top} {\bf f}({\bf a}_i|X)\right),
\end{equation}
where ${\bf f}$ assigns $m$ feature values to ${\bf a}_i$. 
We will discuss the specifics of ${\bf f}(\cdot|\cdot)$ in the next Section. For $S_{i,j}(X)$, we choose the Gaussian kernel:
$S_{i,j}(X) =e^{-\frac{\|{\bf a}_i-{\bf a}_j\|^2}{\sigma^2}}.$
\subsection{Learning DPP model}
\label{subsec::learning::DPP}
We now formulate the learning framework of the conditional DPP as follows. We denote the {\em training set}  as a sequence of ordered pairs ${\cal T}:=(X_1,Y_1),\dots,(X_K,Y_K)$, where $X_k$ is the input and $Y_k\subseteq {\cal Y}(X_k)$ is the output.  Then the learning problem is   the maximization of the log-likelihood of $\cal T$:
\begin{equation}\label{eq::max::log-likelihood}
(\bm{\theta}^*,\sigma^*) = \arg\max\limits_{(\bm{\theta},\sigma)}{\cal L}({\cal T};\bm{\theta},\sigma),
\end{equation}
where ${\cal L}({\cal T};\bm{\theta},\sigma) =$
\begin{equation}\label{eq::log-likelihood}
\scalebox{1.0}{$
 \log\prod\limits_{k=1}^K {\cal P}_{\bm{\theta},\sigma} (Y_k|X_k) = \sum\limits_{k=1}^K\log {\cal P}_{\bm{\theta},\sigma} (Y_k|X_k),$}
\end{equation}
where ${\cal P}_{\bm{\theta},\sigma}\equiv {\cal P}_L$ parameterized  by $\bm{\theta}$ and $\sigma$. 
The reason for choosing the log-linear model for quality measure  and Gaussian kernel  is the fact that under these models, ${\cal L}({\cal T};\bm{\theta},\sigma)$ becomes a concave function of $\bm{\theta}$ and $\sigma$~\cite[Proposition 4.2]{kulesza2012determinantal}. 
\vspace{-.2cm}
\subsection{Inference}
\label{subsec::inference} 
\vspace{-.1cm}
We now estimate $\hat{Y}$ given $X$ using the trained conditional DPP. This phase is known as the {\em testing} or inference phase. In what follows, we present two methods for choosing $\hat{Y}$.  
\subsubsection{Sampling from DPP}
The first option is to draw random sample from the DPP,  i.e., ${\bf Y}\sim {\cal P}_{\bm{\theta}^*,\sigma^*}(\cdot|X)$ and set $\hat{Y} = {\bf Y}$.  We now discuss the sampling scheme for a general DPP which naturally extends to sampling from conditional DPP.  We  start with drawing a random sample from a special class of DPP, known as the {\em elementary} DPP and will use this method to draw samples from a general DPP.  
\begin{algorithm}
\caption{Sampling from a DPP}\label{algo::DPP::sampling}
\begin{algorithmic}[1]
\Procedure{SampleDPP}{$L$}
\State  Eigen decomposition of $L$: $L = \sum_{n=1}^N \lambda_n {\bf v}_n {\bf v}_n^{\top}$
\State $J = \emptyset$
\For{$n=1,\dots,N$}
\State $J\gets J\cup \{n\}$ with probability $\frac{\lambda_n}{\lambda_n+1}$
\EndFor
\State $V\gets \{{\bf v}_n\}_{n\in J}$
\State $Y\gets \emptyset$
\State $B = \begin{bmatrix}{\bf b}_1,\dots,{\bf b}_n\end{bmatrix} \gets V^{\top}$
\For{1 to $|V|$}
\State select $i$ from ${\cal Y}$ with probability $\propto \|{\bf b}_i\|^2$
\State $Y\gets Y\cup\{i\}$
\State ${\bf b}_j\gets {\rm Proj}_{\bot {\bf b}_i} {\bf b}_j$
\EndFor
\Return $Y$
\EndProcedure
\end{algorithmic}
\end{algorithm}
  A DPP on ${\cal Y}$ is called {\em elementary} if every eigenvalue of its marginal kernel lies in $\{0,1\}$. Thus an elementary DPP can be denoted as ${\cal P}^V$ where $V=\{{\bf v}_1,\dots,{\bf v}_k\}$ is the set of $k$ orthonormal vectors such that $K^V = \sum_{{\bf v}\in V}{\bf v}{\bf v}^{\top}$. 
  We now establish that the samples drawn according to ${\cal P}^V$  always have fixed size. 
\begin{lemma}\label{lemm::elementary::DPP::Sample::cardinality} If ${\bf Y}\sim {\cal P}^V$, then $|{\bf Y}| = |V|$ almost surely.
\end{lemma}
\begin{IEEEproof}
If $|{Y}|>|V|$, ${\cal P}^V(Y\subseteq {\bf Y})=0$ since ${\rm rank}(K^V) = |V|$. Hence $|{\bf Y}|\leq |V|$. Now,
$
\E[|{\bf Y}|] = \E[\sum_{n=1}^N {\bf 1}({\bf a}_n\in{\bf Y})]=\E\sum_{n=1}^N[ {\bf 1}({\bf a}_n\in{\bf Y})] = \sum_{n=1}^N K_{n,n} = {\rm trace}(K) = |V|.$
\end{IEEEproof}
Our objective is to find a method to draw a  $k=|V|$ length sample $Y\subseteq{\cal Y}$. Using Lemma~\ref{lemm::elementary::DPP::Sample::cardinality}, ${\cal P}^V(Y) ={\cal P}^V(Y\subseteq {\bf Y}) =  {\rm det}(K_{Y}^V)$. In what follows, we present an iterated sampling scheme that samples  $k$ elements of $Y$  from ${\cal Y}$ without replacement  such that the joint probability of obtaining $Y$ is $ {\rm det}(K_{Y}^V)$.  
Without loss of generality, we assume $Y = \{1,2,\dots,k\}$. 
  Let $B = \begin{bmatrix}
 {\bf v}_1^{\top},\dots,{\bf v}_k^{\top}   
\end{bmatrix}^{\top}$ be the matrix whose rows contain the eigenvectors of $V$. Then, $K^V = BB^{\top}$ and ${\rm det}(K^V_Y) = ({\rm Vol}(\{{\bf b}_i\}_{i\in Y}))^2$, where ${\rm Vol}(\{{\bf b}_i\}_{i\in Y})$ is the   volume of the  parallelepiped spanned by the column vectors (${\bf b}_i$-s) of $B$.  
  Now,  ${\rm Vol}(\{{\bf b}_i\}_{i\in Y}) = \|{\bf b}_1\|{\rm Vol}(\{{\bf b}_i^{(1)}\}_{i=2}^k)$, where 
  ${\bf b}_i^{(1)} = {\rm Proj}_{\bot {\bf b}_1}{\bf b}_i$ denotes the projection of $\{{\bf b}_i\}$ onto the subspace orthogonal to ${\bf b}_1$. Proceeding in the same way, 
\begin{multline}\label{eq::sampling::equation}
{\rm det}(K^V_Y) = ({\rm Vol}(\{{\bf b}_i\}_{i\in Y}))^2 =\\ \|{\bf b}_1\|^2\|{\bf b}_2^{(1)}\|^2\dots \|{\bf b}_k^{(1,\dots,k-1)}\|^2.     
\end{multline}
 Thus, the $j^{th}$ step ($j>1$) of the sampling scheme assuming $y_1 = 1,\dots,y_{j-1}=j-1$ is to select $y_j=j$ with probability proportional to $\|{{\bf b}}_j^{(1,\dots,j-1)}\|^2$ and project $\{{\bf b}_i^{(1,\dots,j-1)}\}$ to the subspace orthogonal to ${\bf b}_j^{(1,\dots,j-1)}$. By \eqref{eq::sampling::equation}, it can be guaranteed that ${\cal P}^V(Y) = {\rm det}(K^{V}_Y)$. 
 
 Having derived the sampling scheme for an elementary DPP, we are in  a position to draw samples from a DPP. The sampling scheme  is enabled by the fact that a DPP can be expressed as a mixture of elementary DPPs. The result is formally stated in the following Lemma.
 \begin{lemma}\label{lemm::DPP::sampling::lemma}
A DPP with kernel $L= \sum_{n=1}^N\lambda_n{\bf v}_n {\bf v}_n^{\top}$ is a mixture of elementary DPPs:
\begin{align}
\scalebox{1.0}{$
{\cal P}_L = \sum\limits_{J\subseteq \{1,\dots,N\} }{\cal P}^{V_J} \prod\limits_{n\in J} \frac{\lambda_n}{1+\lambda_n},$}
\end{align} 
where $V^J = \{{\bf v}_n\}_{n\in J}$. 
\end{lemma} 
\begin{IEEEproof}Please refer to \cite[Lemma~2.6]{kulesza2012determinantal}. 
\end{IEEEproof}
 Thus, given an eigendecomposition of $L$, the DPP sampling algorithm can be separated into  two main steps: (i)  sample an elementary DPP ${\cal P}^{V_J}$ with probability proportional to $\prod_{n\in J}\lambda_n$, and (ii)  sample a sequence of length $|J| $ from the elementary DPP ${\cal P}^{V_J}$. The steps discussed thus  far are summarized in Alg.~\ref{algo::DPP::sampling}.
\subsubsection{MAP inference}
A more formal technique is to obtain the maximum {\em a posteriori} (MAP) set, i.e., $\hat{Y} = \arg\max_{Y\subseteq {\cal Y}(X)}{\cal P}_{\bm{\theta}^*,\sigma^*}(Y|X)$. But, finding $\hat{Y}$ 
is an NP-hard problem because of  the exponential order search space  ${Y\subseteq {\cal Y}(X)}$. However, one can construct computationally efficient  MAP 
inference algorithm which has similar complexity as  random sampling. Due to space limitations, more formal discussions  of these approximation techniques are outside the scope of the paper. We refer to~\cite{NIPS2012_4577} for one possible near-optimal MAP inference scheme for DPPs which will be used  in the numerical simulations. 
\section{Case Study: Link Scheduling}
We will now introduce the link scheduling problem where we will apply the  DPPL discussed in the previous Section. 
\vspace{-1em}
\subsection{System Model}
\label{sec::system::model} 
\begin{figure}
\centering
\includegraphics[scale=0.5,trim={.25cm .25cm .25cm .25cm},clip]{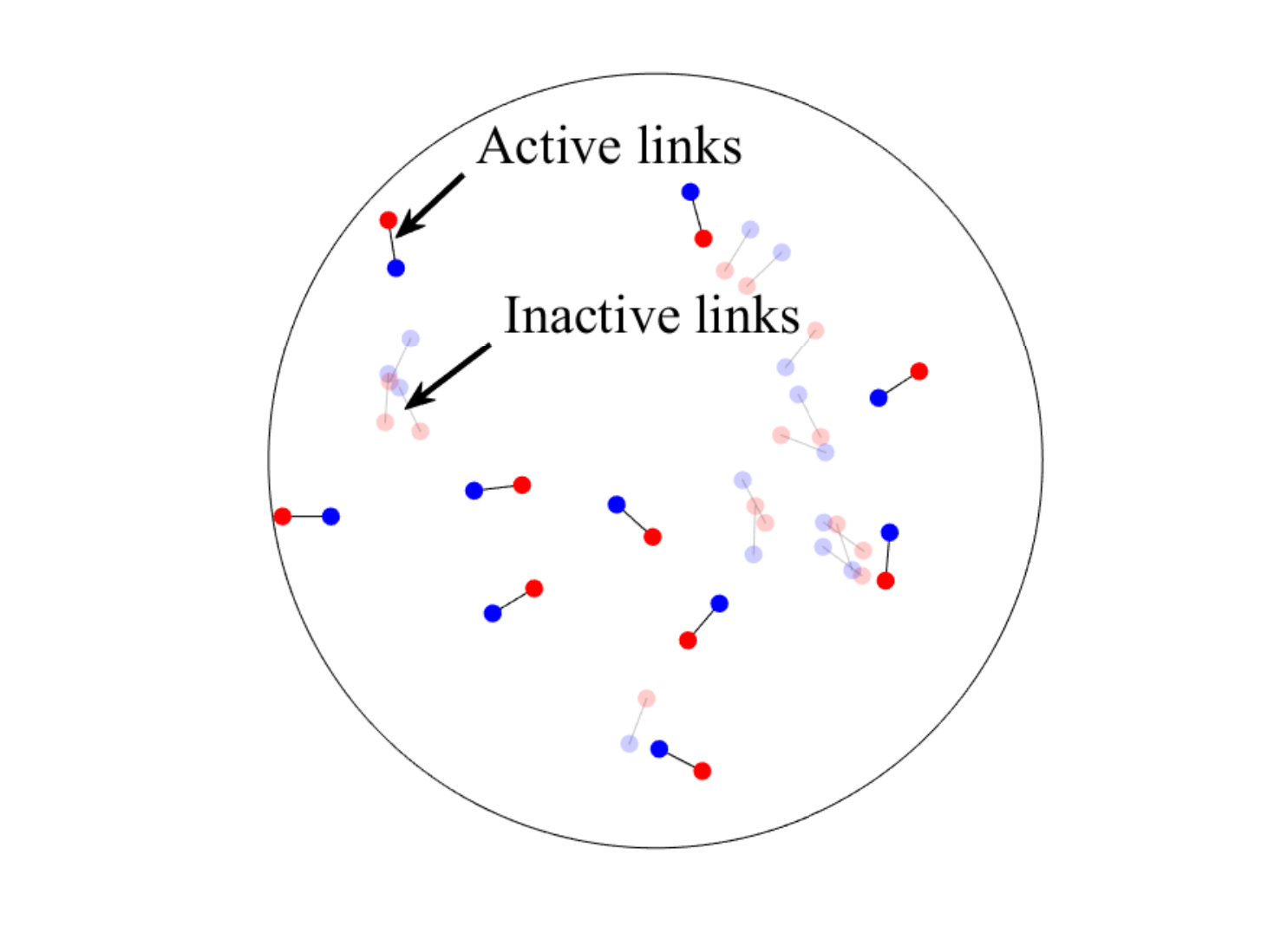}
\caption{Illustration of link scheduling as a subset selection problem. A realization of the network ($M=24$) with the active link subset (${\cal E}^*$).  Details of the network model are mentioned in  Section~\ref{sec::result}.}\label{fig::network::topology}
\end{figure}

We consider a wireless network with  $M$ Tx-Rx  pairs with  fixed link distance $d$. 
The network can be represented as a directed  bipartite graph ${\cal G}:=\{{\cal N}_t,{\cal N}_r,{\cal E}\}$, where ${\cal N}_t$ and ${\cal N}_r$ are the independent sets of vertices denoting the set  of Tx-s and Rx-s, respectively and ${\cal E}:=\{(t,r)\}$ is the set of directed edges  where  $t\in {\cal N}_t$ and $r\in {\cal N}_r$. Since each Tx has its dedicated Rx, the in-degree and out-degree of each node in ${\cal N}_t$ and ${\cal N}_r$ are one. 
Also $|{\cal N}_t| =|{\cal N}_r| =  |{\cal E}| =M$.   An illustration  of the network topology is presented in Fig~\ref{fig::network::topology}. {Let ${\cal K}_{{\cal N}_t,{\cal N}_r}^{\cal W}$ be the complete weighted bipartite graph on ${\cal N}_t,{\cal N}_r$ with ${\cal W}(i,j) = \g_{ij}$ for all $i\in{\cal N}_t,j\in {\cal N}_r$. Here $\g_{ij}$ denotes the channel gain between Tx $i$ and Rx $j$.} 
\subsection{Problem Formulation}
We assume that each link can be either in {\em active} or {\em inactive} state. A link is active when the  Tx  transmits at a power level $p_{\rm h}$ and is inactive when the Tx transmits at a power level $p_{\ell}$ (with $0\leq p_{\ell}<p_{\rm h}$).  Each link transmits over the same frequency band whose bandwidth is assumed to be unity. Then the sum-rate on the $l^{th}$ link is given by $\log_2\left(1+\gamma_l\right)$, where $\gamma_l$ is the $\sinr$ at the $l^{th}$ Rx:  $\gamma_l = \frac{\g_{ll} p_l}{\sigma^2+\sum
_{e_j\in
     {\cal E}}^{j\neq l}\g_{jl}p_j}$. Here $\sigma^2$ is thermal noise power.   
The sum-rate maximization problem can be expressed as follows. 
\begin{subequations}\label{eq::sum::rate::maximize}
\begin{alignat}{2}
&\text{maximize } \sum\limits_{e_l\in{\cal E}}\log_2\left(1+\gamma_l\right),\\
&\text{subjected to } p_l\in\{p_{\ell},p_{\rm h}\}\label{eq::constraint::integer},
\end{alignat}
\end{subequations}
where the variables are $\{p_l\}_{e_l\in{\cal E}}$. An optimal subset of simultaneously active links denoted as ${\cal E}^*\subseteq{\cal E}$ is the solution of \eqref{eq::constraint::integer}. Thus, $p_l = p_{\rm h},\ \forall\ {e_l}\in{\cal E}^*$ and $p_l = p_{\ell}, \ \forall\ {e_l}\in {\cal E}\setminus{\cal E}^*$.
\subsection{Optimal Solution}
The optimization problem in \eqref{eq::sum::rate::maximize} is NP hard~\cite{weeraddana2012weighted}. However, for bipartitle networks the problem can be solved by a low-complexity heuristic algorithm based on GP (see Alg.~\ref{algo::successive::approximation}). For completeness, we have provided the rationale behind its formulation in Appendix~\ref{app::construction}.  For further details on solving  the  general class of link scheduling problems, the reader is referred to~\cite{weeraddana2012weighted}. Fig.~\ref{fig::network::topology} demonstrates a realization of the network and ${\cal E}^*$ chosen by Alg.~\ref{algo::successive::approximation}.
\begin{algorithm}
\caption{Optimization algorithm for \eqref{eq::sum::rate::maximize}}\label{algo::successive::approximation}
\begin{algorithmic}[1]
\Procedure{SumRateMax}{${\cal K}^{\cal W}_{\cal N},\cal E$}
\State Initialization: given tolerance $\epsilon>0$, set ${\bf P}_0=
\{p_{l,0}\}$. Set $i=1$. Compute the initial $\sinr$ guess $\hat{\bf \gamma}^{(i)}=\{\gamma_{l}^{(i)}\}$. 
\Repeat
\State Solve the GP:
\begin{subequations}\label{eq::GP::formulation}
\begin{alignat}{2}
\text{minimize } &K^{(i)}\prod \gamma_{l}^{-\frac{\hat{\gamma}_l^{(i)}}{1+\hat{\gamma}_l^{(i)}}}\\
\text{subject to }&\beta^{-1}\hat{\gamma}_l^{(i)}\leq \gamma_l \leq \beta \hat{\gamma}_{l}^{(i)}, {e_l}\in{\cal E},\label{eq::modified::2::constaint}\\
&\sigma^2 \g_{ll}^{-1}p_{l}^{-1}\gamma_{l}+\sum\limits_{j\neq l}\g_{ll}^{-1}\g_{jl}p_jp_l^{-1}\gamma_{l}\leq 1,e_l\in{\cal E},\\
& p_l\leq p_{\max},\ \forall\ e_l\in{\cal E}.  
\end{alignat}
\end{subequations}
with the variables $\{p_l,\gamma_l\}_{e_l\in{\cal E}}$. Denote the solution by $\{p_l^*,\gamma_l^*\}_{e_l\in{\cal E}}$. 
\Until{$\max_{e_l\in{\cal E}}|\gamma_{l}^*-\hat{\gamma}_l^{(i)}|\leq \epsilon$}  
\If{$p_l\geq p_{\rm th}$}
     \State $p_l = p_{\rm h}$
\Else
     \State $p_l = p_{\ell}$
\EndIf
\Return ${\cal E}^*$
\EndProcedure
\end{algorithmic}
\end{algorithm}
\subsection{Estimation of  optimal subset with DPPL}
\begin{figure}
\centering
\includegraphics[scale=0.45]{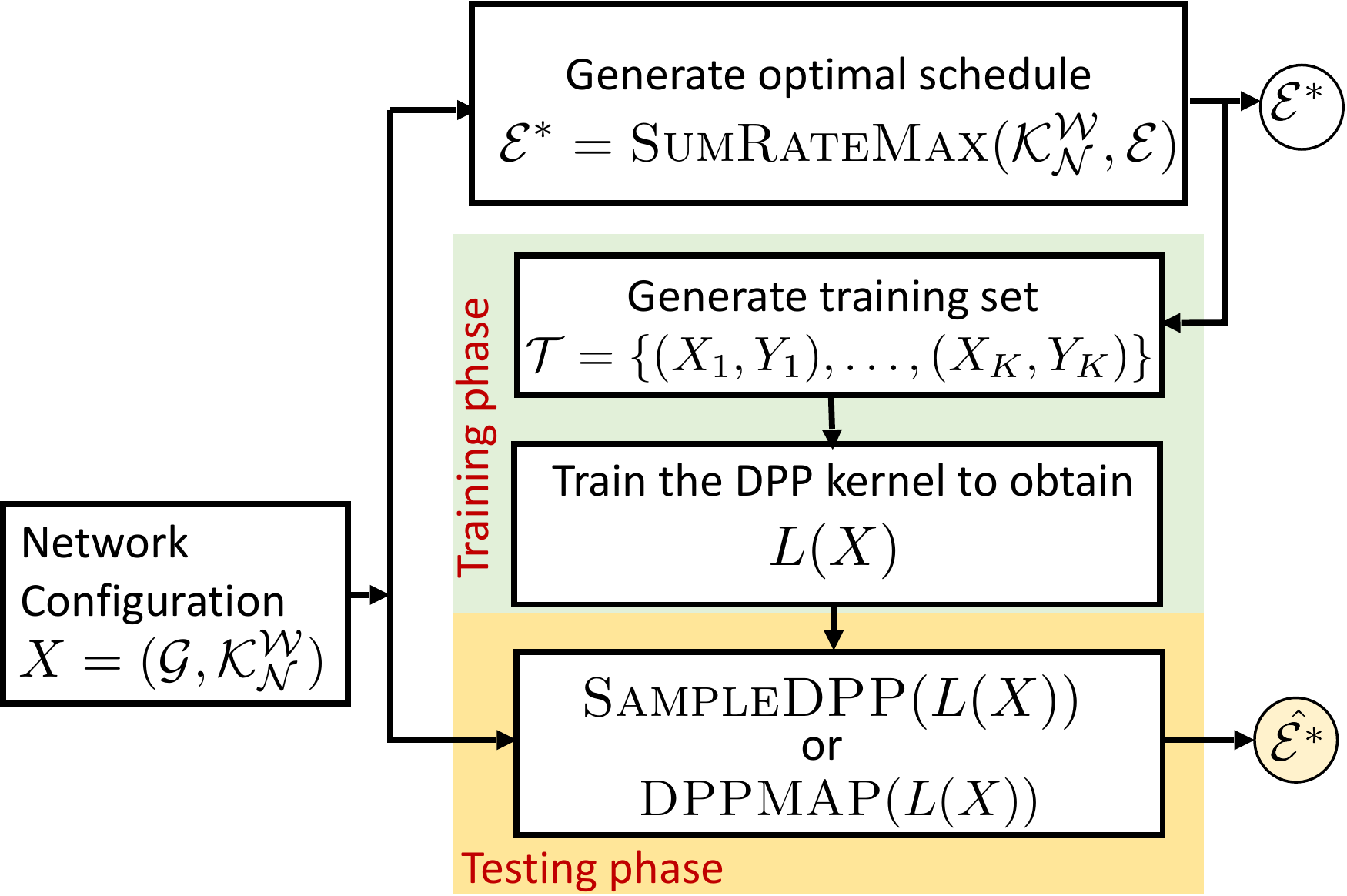}
\caption{Block diagram of  DPPL  for the link scheduling problem.}\label{fig::blockdiagram}
\end{figure}

We will now model the problem of optimal subset selection ${\cal E}^*\subseteq {\cal E}$  with DPPL.  
We train the DPP with a sequence of networks and the optimal subsets obtained by Alg.~\ref{algo::successive::approximation}. For the training phase, we set $X_k = ({\cal K}_{{\cal N}_t,{\cal N}_r}^{\cal W},{\cal E},{\cal E}^*)_k$ as the $k^{th}$ realization of the network and its optimal subset. The quality and diversity measures are set as: 
$g({\bf a}_i|X) := \exp\left(\theta_1 \zeta_{ll} p_{\rm h} +\theta_2 I_1 + \theta_3 I_2\right),
$
where $I_1 = p_{\rm h}\zeta_{j'i}$ with ${j}' = \arg\max_{j=1,\dots,L\neq i}\{\zeta_{ji}\}$ and $I_2 = p_{\rm h}\zeta_{{j}''i}$ with  ${j}'' = \arg\max_{j=1,\dots,L\neq i,j'}\{\zeta_{ji}\}$ are the two strongest interfering powers, and
$
S_{i,j}(X) =\exp{-(\|{\bf x}(t_i)-{\bf x}(r_j)\|^2 + \|{\bf x}(t_j)-{\bf x}(r_i)\|^2)/\sigma^2},
$
where ${\bf x}({t_i})$ and ${\bf x}({r_j})$ denote the locations of  Tx $t_i\in{\cal N}_t$ and Rx $r_j\in{\cal N}_r$, respectively. The ground set of the DPP ${\cal Y}(X) = {\cal E}$. We denote the subset estimated by  DPPL in the testing phase as $\hat{{\cal E}^*}$. The block diagram of the DPPL is illustrated in Fig.~\ref{fig::blockdiagram}. In order to ensure the reproducibility of the results, we provide the  Matlab implementation of the   DPPL for this case study in~\cite{SahaDPPLCode}.
\subsection{Results and Discussions}\label{sec::result}
We now demonstrate the performance of DPPL through numerical simulations. 
We construct the network by distributing $M$ links with  $d=1$ m  within a disc of radius $10$ m uniformly at random. We assume channel gain is dominated by the power law path loss, i.e., $\zeta_{ij} = \|{\bf x}(t_i)-{\bf x}(r_j)\|^{-\alpha}$, where $t_i\in{\cal N}_t$, $r_j\in{\cal N}_r$, and $\alpha=2$ is the pathloss exponent. The network during training and testing phases  was generated by  setting $M\sim{\tt Poisson}(\bar{M})$ with  $\bar{M}= 20$. The instances where $M=0$ were discarded. We set $p_{\rm h}/\sigma^2= p_{\max}/\sigma^2= 33$ dB, $p_{\ell}/\sigma^2=13$ dB,  and $p_{\rm th}=23$ dB. The training set ${\cal T}$ was constructed by $K = 200$ independent realizations of the network. Note that changing $K$ from $20$ to $200$ did not change the values of $\sigma^*$ and ${\bm{\theta}}^*$ ({$\sigma^* = 0.266, {\bm{\theta}}^* = \begin{bmatrix}
996,675,593
\end{bmatrix}$}) significantly.  
In Fig.~\ref{fig::main-result::sum::rate}, we plot the empirical cumulative distribution functions (CDFs) of the sum-rates obtained by Alg.~\ref{algo::successive::approximation} and DPPL. We observe that the sum-rate obtained by  DPPL framework closely approximates the $\max$-sum-rate. We also notice that DPP MAP inference gives better sum-rate estimates than DPP sampling. 
 We further compare the performance  with the well-known SG-based  model where the simultaneously active links are modeled as {\em independent thinning} of the actual network~\cite{blaszczyszyn2018determinantal}. In particular, each link is assigned $p_{\rm h}$ according to an independent and identically distributed (i.i.d.) Bernoulli random variable with probability $\xi$. {We estimate $\xi$ by averaging the activation of a randomly selected link which is equivalent to: $\xi = \sum_{k=1}^K {\bf 1}(e_i\in {\cal E}_k^*)/K$ for a fixed $i$}. We see that the sum-rate under independent thinning is significantly lower than the one predicted by DPP. The reason is the fact that the independent thinning scheme is not rich enough to capture  spatial repulsion which  exists across the links of ${\cal E}^*$. 
\subsubsection{Run-time Comparison}
\begin{figure*}
  \begin{minipage}{\linewidth}
      \centering
      \begin{minipage}{0.3\linewidth}
          \begin{figure}[H]
              \includegraphics[width=.981\linewidth]{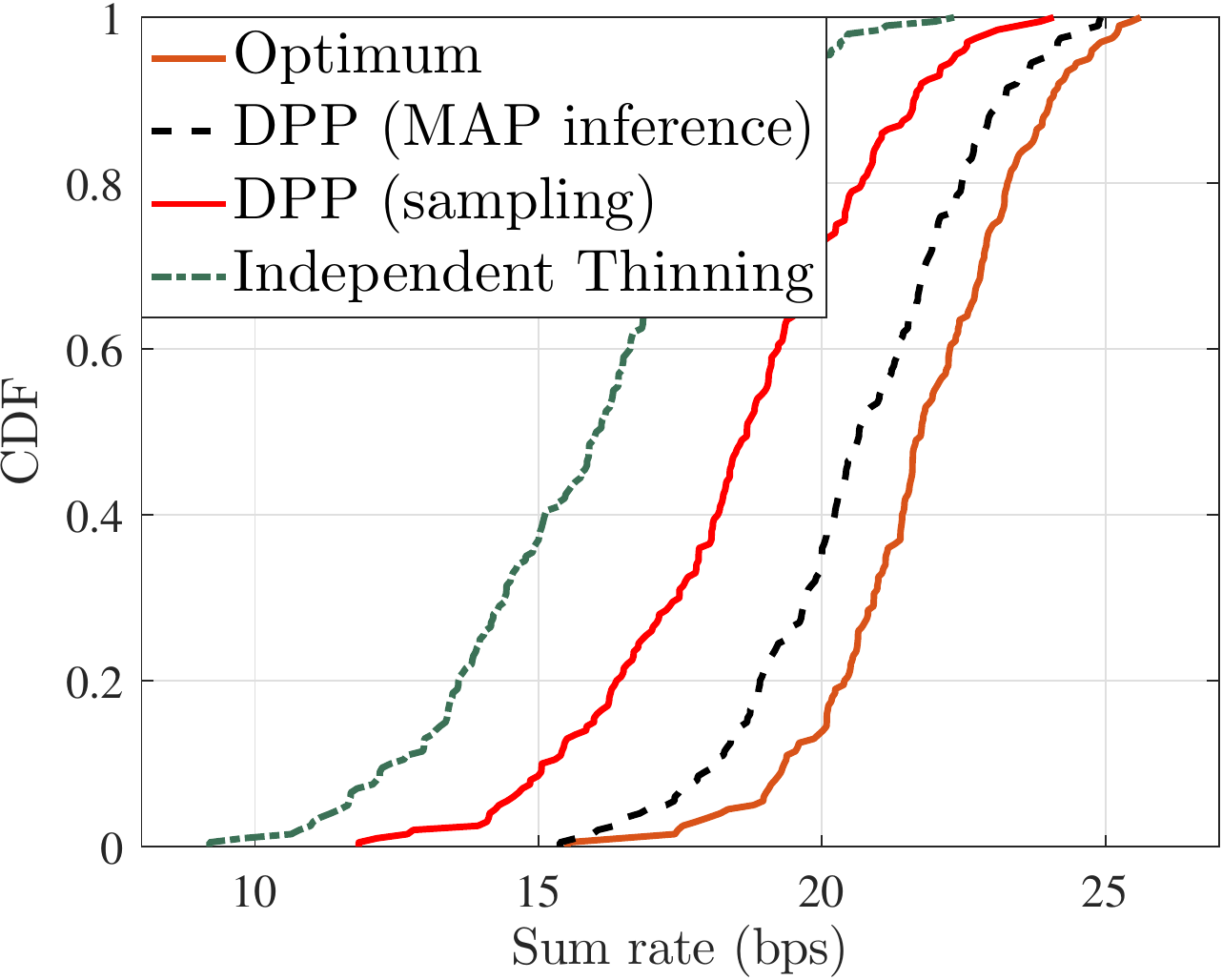}
\caption{CDF of sum-rate obtained by different subset selection schemes.}\label{fig::main-result::sum::rate}
          \end{figure}
      \end{minipage}
      \begin{minipage}{0.3\linewidth}
          \begin{figure}[H]
              \includegraphics[width=\linewidth]{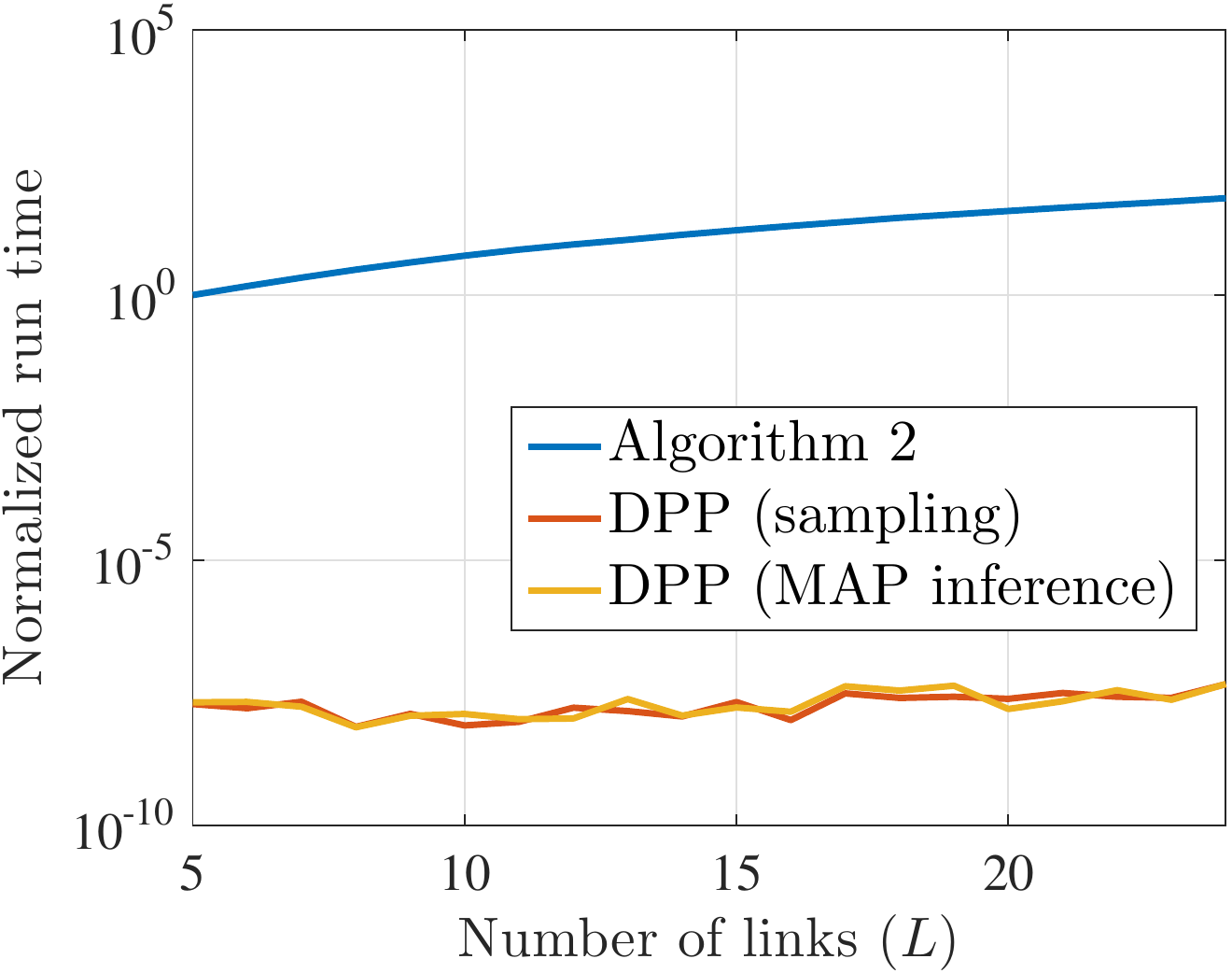}
\caption{Comparison of run-times of Alg.~\ref{algo::successive::approximation} and DPPL in testing phase.}\label{fig::run::time}
          \end{figure}
      \end{minipage}
            \begin{minipage}{0.3\linewidth}
          \begin{figure}[H]
              \includegraphics[width=\linewidth]{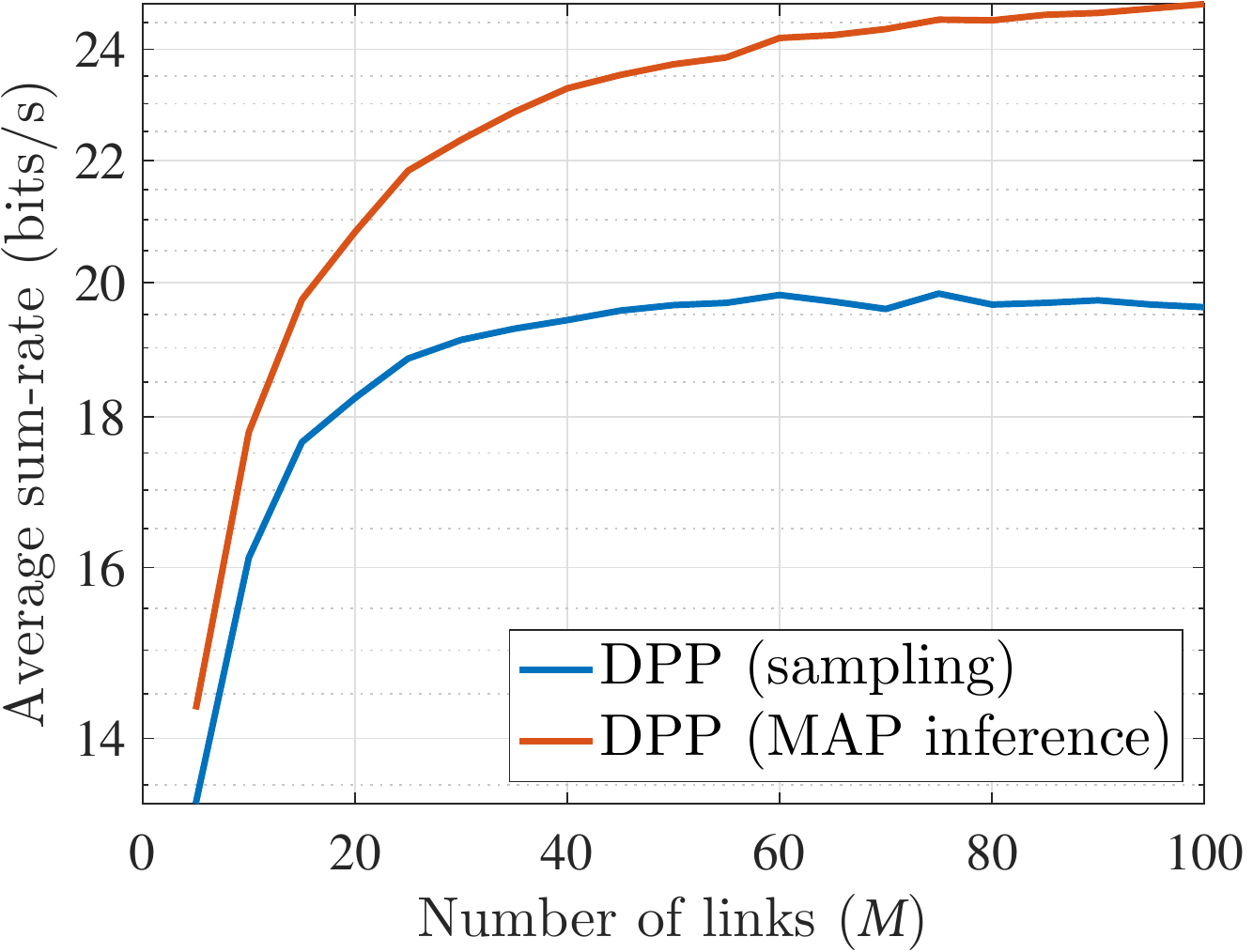}
              \caption{Average rates obtained for different network sizes using DPPL.}\label{fig::average::rate::compare}
          \end{figure}
      \end{minipage}
  \end{minipage}
\end{figure*}

Another key strength of the proposed  DPPL  appears when we compare its run-time  in the testing phase and Alg.~\ref{algo::successive::approximation} applied on a network $({\cal K}_{{\cal N}_t,{\cal N}_r}^{\cal W},{\cal E})$. In Fig.~\ref{fig::run::time}, we  plot the run-times of different subset selection schemes for different network sizes. The absolute values of run-times were obtained averaging the run-times of  all the schemes over $1000$ iterations in the same  computation environment.  In order to obtain a unit-free measure, we  normalize these absolute values by dividing them with the average absolute run-time of Alg.~\ref{algo::successive::approximation} for $M=5$. 
We observe that DPPL is at least $10^5$ times faster than Alg.~\ref{algo::successive::approximation}.  The run-time of Alg.~\ref{algo::successive::approximation}   increases exponentially with $M$ whereas  run-times of the DPPL  scale as {some polynomial order} of $M$.    

{Note that DPPL is not just a sum-rate estimator of the network, but it estimates the optimal subset  of links  $\hat{{\cal E}^*}$ significantly faster than the optimization algorithms. Thus,  DPPL  can be implemented in real networks  to determine ${\cal E}^*$ even when the network size is large.}   In Fig.~\ref{fig::average::rate::compare}, we plot the sum-rates averaged over $10^3$ network realizations for a given value of $M$. Note that evaluating $\max$-sum-rates for higher values of $M$ using Alg.~\ref{algo::successive::approximation} is nearly impossible due to its exponentially increasing run-time. Quite interestingly,  {DPPL}, thanks to its fast computation, provides some  crisp insights on the network behavior: as more number of links are added, the estimated $\max$-sum-rate tends to saturate (see Fig.~\ref{fig::average::rate::compare}). {This is expected because as long as the resources are fixed, there will be a limit on the number of simultaneously active links (irrespective of $M$) that would maximize the sum-rate. If the number of active links  is more than this limit, sum-rate may decrease because of the increased interference.}    Also we observe that the performance difference between  MAP-inference and DPP-sampling increases significantly at higher values of $M$. 
\vspace{-.1cm}
\section{Conclusion} {
In this paper, we identified a general class of subset selection problems in wireless networks which can be solved by jointly leveraging ML and SG, two fundamentally different mathematical  tools used in communications and networking. To solve these problems, we developed the  DPPL framework, where the DPP orginiates from SG and its learning applications have been fine-tuned by the ML community. When applied to a special case of wireless link scheduling, we found that the DPP is able to {\em learn} the underlying  quality-diversity tradeoff of the optimal subsets of simultaneously active links.  This work has numerous extensions. From the SG perspective, it is of interest to compute analytical expressions of the key performance metrics of the network such as the mean interference at  a typical receiver or the average rate by leveraging the analytical tractability of DPPs.  From the ML perspective, the DPPL can be extended to include time as another dimension and solve the  subset selection problems over time (e.g. the scheduling problems in cellular networks, such as the proportional fair scheduling) using the space-time version of the  DPPL  (also known as the {\em dynamic DPP}~\cite{Osogami2018DynamicDP}). From the application side, this framework can be used to solve other subset selection problems such as the user group selection in a downlink multiuser  multiple-input-multiple-output (MIMO) setting.}
\appendix
\subsection{Formulation of Alg.~\ref{algo::successive::approximation}}
\label{app::construction}
  Since \eqref{eq::sum::rate::maximize} is an integer programming problem, the first step is to solve the relaxed version of the problem assuming continuous power allocations. In particular, we modify the  integer constraint \eqref{eq::constraint::integer} as  $0\leq p_l\leq p_{\max}$. 
Since $\log_2(\cdot)$ is an increasing function, the problem can be restated as:
\begin{subequations}\label{eq::problm::formulate::intermediate}
\begin{alignat}{3}\label{eq::problm::formulate::intermediate::objective}
&\min_{\{p_l\}_{e_l\in{\cal E}}} &&\scalebox{0.9}{$\prod\limits_{e_l\in{\cal E}}(1+\gamma_{l})^{-1}$}\\
&\text{s.t.}&&\gamma_{l} = \frac{\g_ll p_l}{\sigma^2+\sum_{j\neq l}\g_{jl}p_{jl}}, e_l \in{\cal E}\label{eq::equality::constraint}\\
& &&0\leq p_{l}\leq p_{\max}\ \forall\ e_l\in{\cal E}. 
\end{alignat}
\end{subequations}
Since the objective function  is
decreasing in  $\gamma_l$, we can replace the equality in \eqref{eq::equality::constraint}  with inequality. Using the auxiliary variables $v_l\leq 1+\gamma_l$,  \eqref{eq::problm::formulate::intermediate} can be formulated as:
\begin{subequations}
\begin{alignat}{3}
&\min_{\{p_l,\gamma_l,v_l\}} \prod_{e_l\in{\cal E}}v_l^{-1}\label{eq::objective::monomial}\\
& \text{s.t. } v_l\leq 1+\gamma_l,\ \forall e_l\in{\cal E} \label{eq::posynomial}\\
& \sigma^2 \g_{ll}^{-1}p_l^{-1}\gamma_l + \sum_{j\neq l}\g_{ll}^{-1}\g_{jl}p_{j}p_l^{-1}\gamma_l\leq 1, e_l\in{\cal E},\label{eq::third::constraint}\\
& 0\leq p_l p_{\max}^{-1}\leq 1.
\end{alignat}\label{eq::CGP}
\end{subequations}
Now in  \eqref{eq::CGP}, we observe that \eqref{eq::objective::monomial} is a monomial function, \eqref{eq::posynomial} contains posynomial function in the right hand side (RHS), and all the constraints contain either monomial or posynomial functions. Hence, \eqref{eq::CGP} is a complementary GP~\cite{Boyd2007}. If the posynomial in \eqref{eq::posynomial} can be replaced by a monomial, \eqref{eq::CGP} will be a standard GP. Since GPs can be reformulated as convex optimization problems, they can be solved efficiently irrespective of the scale of the problem. One way of approximating \eqref{eq::CGP} with a GP at a given point $\{\gamma_l\}=\{\hat{\gamma_l}\}$ is to replace the posynomial $1+{\gamma}_l$ by a monomial $k_l{{\gamma}}_l^{\alpha_l}$.  
 From $1+\hat{\gamma}_l =k_l{\hat{\gamma}}_l^{\alpha_l} $, we get 
\begin{equation}\label{eq::k::and::alpha} 
\scalebox{1.0}{$
 \alpha_l= \hat{\gamma}_l(1+\hat{\gamma}_l)^{-1},\quad k_l = {\hat{\gamma}_l}^{-\alpha_l}(1+\hat{\gamma}_l).$}
  \end{equation} 
 Also note that $1+\gamma_l\geq k_l\gamma_l^{\alpha_l}$, $\forall\ \gamma_l>0$ for $k_l>0$ and $0<\alpha_l<1$. Thus the local  approximation of \eqref{eq::CGP} will still satisfy the original constraint \eqref{eq::posynomial}. The modified inequality constraint becomes 
 \begin{equation}\label{eq::modified::constaint}
 \scalebox{1.0}{$
 v_l\leq k_l\gamma_l^{\alpha_l},\ \forall\ e_l\in{\cal E},$}
 \end{equation}
 where $k_l$ and $\alpha_l$ are obtained by \eqref{eq::k::and::alpha}.
 
 Since \eqref{eq::objective::monomial} is a decreasing function of $v_l$, we can substitute $v_l$ with its maximum value $k_l\gamma_l^{\alpha_l}$, which satisfies the other inequality constraints. Thus, $v_l$ can be eliminated as:
 \begin{equation}
 \scalebox{1.0}{$
 \prod_{e_l\in{\cal E}}v_l^{-1} = \prod_{e_l\in{\cal E}}k_l^{-1}\gamma_l^{-\alpha_l} = K\prod_{e_l\in{\cal E}}\gamma_l^{-\frac{\hat{\gamma}_l}{1+\hat{\gamma}}_l}, $}
 \end{equation}
 where $K$ is some constant which does not affect the minimization problem. Thus, the $i^{th}$ iteration of the heuristic runs as follows. Let ${\hat{\gamma}_l^{(i)}}$ be the current guess of $\sinr$ values. The GP will provide a better solution ${\hat{\gamma}}_l^{*}$ around the current guess which is set as the initial guess in the next iteration, i.e., ${\hat{\gamma}}_l^{(i+1)}={\gamma}_l^{*}$ unless a termination criterion is satisfied. These steps are summarized in  Alg.~\ref{algo::successive::approximation}.   To ensure that the GP does not drift away from the initial guess  ${\hat{\gamma}_l^{(i)}}$,  a new constraint  \eqref{eq::modified::2::constaint} is added so that $\gamma_l$ remains in the local neighborhood of $\gamma_l^{(i)}$.  Here $\beta>1$ is the control parameter. Smaller the value of $\beta$,  higher is the accuracy of the monomial approximation, but slower is the convergence speed. 
  For a reasonable tradeoff between accuracy and speed, $\beta$ is set to $1.1$. The algorithm terminates with the quantization step which assigns discrete power levels $p_{\ell}$ and $p_{\rm h}$. 
  Once we obtain the optimal power allocation $p_{l}^{*}\in[0,p_{\max}]$, we quantize it into two quantization levels $p_{\ell}$ and $p_{\rm h}$ by setting $p_{l}^{*}=p_{\ell}$ whenever its value lies below some threshold level $p_{\rm th}$ or otherwise $p_{l}^{*}=p_{\rm h}$.
  \vspace{-.5em}
\bibliographystyle{IEEEtran}
\bibliography{DPPProblemFormulationv3.bbl}
\end{document}

%% file: notationITA.tex
\let\emptyset\varnothing
\def\g{{\zeta}}

\def\nb0{{\mathbf{0}}}
\def\nb1{{\mathbf{1}}}

\newtheorem{lemma}{Lemma}

\newtheorem{remark}{Remark}

\def\E{\mathbb{E}}

\def\R{\mathbb{R}}

\def\sinr{\mathtt{SINR}}			

\def\matern{Mat\'ern\ }

